\documentclass[acmtocl]{acmtrans2m}

\acmVolume{1}
\acmNumber{1}
\acmYear{06}
\acmMonth{June}

\newcommand{\BibTeX}{{\rm B\kern-.05em{\sc i\kern-.025em b}\kern-.08em
    T\kern-.1667em\lower.7ex\hbox{E}\kern-.125emX}}

\title{Use MPLS in LANs 
(Enhancement performance and safety)}
\author{Atos Ramos Alves}
\begin{abstract} 
To demonstrate the result of researches in laboratory with the focus in exhibiting the Real impact of the use of the technology MPLS in LAN. Through these researches we will verify that the investment in this technology is shown, of the point of view cost/benefit, very interesting, being necessary, however, the adoption of another measured, in order to settle down a satisfactory level in the items Quality and safety in the sending of packages in VPN but assisting to the requirement latency of the net very well being shown in the tests that it consumes on average one Tuesday leaves of the time spend for the same function in routing IP. 
\end{abstract}
\category{k.6.3}{Management of Computing and Information Systems}{Software Management}%
[software selection]




\begin{document}

\begin{bottomstuff} 
Authors' addresses: Atos Ramos Alves, Docents, University of Managerial Sciences UNA, Brazil-Minas Gerais-Belo horizonte, 30570-310
\end{bottomstuff}

\maketitle

\section{Introduction}
This work intends to tell the practical use in the companies and the main characteristics of the technologies VPN (Virtual private network), QOS (Quality Of Service) added to the routing MPLS (MultiProtocolo of keying of labels) as form of fast direction of packages in nets LAN. 
We will present the main results obtained with the simulations using the tools that will be described the front. 
\bibliographystyle{acmtrans}

Sobral (2005) he wrote that MPLS (RFC 3031), technology in the which they are based all the tests of this article, is a technology which directs packages of a computer the other guided through labels (labels) printings in the headers IP. So that there is the understanding of these labels is necessary that the gateways (some models of gateways or even machines turning Linux) involved in the transmission get to recognize them. That technology can be framed in the layer 2,5 of the model OSI, in other words, it is between the connection layer and the one of net, that happens due the header IP it was not projected to support those labels, consequently he doesn't possess available space for a new field needing an addition (extension) of the header MPLS. The existent label in the header MPLS of the packages that are moving are seen as indexes in tables, determining the road to be traveled to reach the next knot of the net. The direction of packages raisin the being a research in extracted table of the packages, differently of the traditional routing, which researches in routing tables, obtaining of the packages just the destiny IP address. With four fields, the header MPLS possesses the field label (the label where the index is contained) as being his main field, we have the field of QoS (Quality of Service), TTL and the interesting field S, which is related with the piling up and hierarchies of nets. In case the field S has value zero, the package will be discarded avoiding to enter in infinite loop, that is a consistence of the use of MPLS for instance in nets of great load. 
Which the importance and reason in adopting a new technology where their direction algorithms don't offer expressive superior speeds to the routing IP, which already a delegated knowledge is? 
Simple, we have other benefits to appeal, among them we can mention: 
- Effective intersection among the universes IP and ATM; 
- Base for VPN's and studies on engineering of traffic; 
- Desegregation between routing and direction; 

In the linux we have an interesting project, a patch for the kernel, with the purpose of providing support MPLS recognizing the existent labels in the packages, obtaining the capacity directing to the next host. The project can be found in http://mpls-linux.sourceforge.net. In this page we can accomplish download on some other packages, as iproute2 customized for instance, capable to identify that protocol and to treat him in the same way as he does today with the headers IP. Other packages can be lowered, as for instance a version of the kernel, iptables, PPP, among other necessary ones for the implantation of MPLS in Linux, all these packages are striped in the own site. A great applicability to the protocol MPLS is in the use of VoIP with MPLS-YOU (Engineering of Traffic), because in this case it is won in performance and quality of voice transmission, since the engineering of traffic helps to solve problems, as for instance loss of packages and other problems that appear when the engineering is not present. The engineering of traffic seeks to provide the best roads for the packages travel, optimizing the time of course reducing collisions, for instance, determining the best route to reach the other host. 
It is really worthwhile to study that protocol, because with him it is day by day possible to improve it in the atmospheres of net of companies of all the loads, reduction of costs, in the acquisition of gateways, for instance.
The use of the technologies VPN, QOS and added MPLS provokes reduction in the latency of the nets LAN (net interns) or WAN (Internet) making possible use of technologies as VoIP, Multimedia and database synchronism (DRBD), same when MPLS is implemented on PCs with System Linux, contradicting found documents, for instance, in \"I Study Experimental of the Technology MPLS: acting Evaluation, service Quality and Engineering of Traffic\" accomplished by NUCARD of the Federal University of Santa Catarina http://www.rnp.br/\_arquivo/wrnp2/2003/eetma01a.pdf
A VPN (virtual net) it is a private tunnel inside of a shared net or public, like the own Internet, for instance and LAN. VPN for if they turn safer they use cryptography techniques of data. Today we have three types more used: PPTP (I record tunnel Point-to-point), IPSEC (Protocol of Safety of the Internet), L2TP (I record Tunnel Layer 2). We will use OpenVPN to implement the tunnels VPN in L2TP with OpenSSL, that they will be used in our tests, for they possess characteristics that will be detailed to proceed. 
OpenVPN, as informed in his/her e-mail address http://openvpn.net/howto.html, is a tool of VPN totally implemented tops of the technology SSL implemented in the levels 2 or 3 of the model OSI. Using extensions of safety of net standard industry records SSL/TTL, it supports methods of users' flexible authentication based on certificates, smart cards, and/or you accredit based in user and password, it also allows user's rules and group, she also can to apply firewall rules in the virtual interface of VPN. Another characteristic of OpenVPN is the compacting of the data after the cryptography generating an increase of data moved in the net a second. This compensates in a lot the time lost with the cryptography. For us to still increase more the delivery speed will explain in the next paragraph on QOS. 
As we can see in RFC2386 the concept of QOS it is a routing mechanism that bases on previously knowing the road to be traveled by the packages. In tool TC will detail more the theme. 
All the tests were accomplished at the laboratories of the company I. Maid in 1985 the company I he assists today to 200 autarchies in municipal districts in Brazil in the states of São Paulo, Minas Gerais and Rio de Janeiro in the areas of Public accounting, Taxation, Control of Materials and Payroll. Tends as main activity the technician-accounting consultantship of municipal districts, what explains the need of high level of safety in the transfer of data, be in LAN (net interns) or in the atmosphere WAN (Internet). Being distributed by 17 floors of a building verified the need to obtain means of activating the traffic in the net LAN without terms that to invest in installation of optic fibers.

\section{METHODOLOGY}

\subsection{PLATFORM OF TESTS}
We created a testbed, or be a controlled atmosphere and conditioned. In this atmosphere we can: (1) to maintain updated databases among two database servants, (2)send of medias continue as audio and video,(3) to do use of tools of tests with net latency, band passer-by, load with ping, latency of protocols and comparison between routing table and routing MPLS. 
All these tools were conditioned in a LiveCD based on System Linux (KNOPPIX). 
LiveCD is a media unit CD, whose characteristic is to allow the initialization (boot) of the equipment without the need of installation of the Operating system in the hard disk. The intention of such scenery was to maintain an atmosphere the stables possible during everyday that they were necessary for accomplishment of the tests. 
To begin each test the equipments they were turned off, in order to maintain the structure the most faithful possible to the characteristics initials of the Operating system, in other words, memoirs SWAP empty, registrars and sharpeners of the Microprocessors and of the clean operating system (reduced to zero). This way we looked for to obtain more reliable results and susceptible to repetition at any moment, with despicable degree of alteration. 
We will present in this section the structure and architecture of the platform of tests set up and used during this project.

\subsection{THE ARCHITECTURE OF THE PLATFORM OF TESTS }
Mounted atmosphere with use of eleven PC's, each one using the System Linux, executed starting from CD. 
The version of the kernel used is 2.6.9, that he brings in a native way the modules of QOS, and supports to the modules of VPN and MPLS that were implemented. Nine PC's possessed two net interfaces, making possible act as gateways in the test atmosphere. The other two PC's acted as origin and destiny of traffic, because they possessed only a net interface. 
We established among the machines a PVC (Virtual Permanet Circuit) limited 220Kbit and latency of 50ms, through the use of QOS. This way we got through CLIP (classic IP), to configure the atmosphere to simulate several local area networks interconnected through nine PC's that will act as gateways in a WAN. In a second moment we removed the limitation created with QOS for us to make the same tests in atmosphere LAN. 
Scripts SHELLS were created in order to execute programs for the control and monitoring of the I traffic, also creating disciplines of lines, classes and filters. 
For exhibition of the dynamics accomplished in the tests, we will call ours conspire (PC's) of: A,B and C. Considering that B symbolizes a group of nine conspires with routing function. 
Imagining that they are in it serializes, it plans her is HER more the left, B is to the center and C more the right. In the it plans B were linked two cables, being one in direction of the it plans and another tied her it plans C. 
He plans her carried out it the measurement functions and generation of the traffic, demarcation and changing of the packages belonging to the traffic EF (Expedited Express) or BE (Best effort), in her the first policing was also accomplished. In this it conspires was executed a script that transformed it in a border gateway (gateway that interconnects a LAN to other gateway). 
In the it plans B implemented scripts that did with that she was capable to act as internal gateway (it colors router) to our net. In the net interfaces it happened the policing of the traffic through the field TOS (Type of Service) contained in the header of the packages. 
It plans it C played destiny part and, also, generating of traffic, once it returned the data (it echoed) for it plans it A. The script to transform it in border gateway, it was also executed in this conspires. 
In spite of the atmosphere servant to be very simple, in him was possible to implement all the tests of VPN and QOS on MPLS that we considered necessary for the development of this article. However we could have several servants, customers, switchs and gateways enriching like this our scenery without any alteration in the structures of tests, being enough small alterations in the configuration files. 
To proceed we will present the tools used for generation and monitoring of the traffic and the tool used to configure the net interfaces in order to if to obtain a service quality for certain classes of services and implementation of VPN.

\subsection{MPLS-LINUX }
MPLS for Linux began as an analyzer of protocols for LDP. It used a package that codifies and it decodes the functions developed by Nortel NET. It was developed originally for N+I Las Vegas iLab' 99 MPLS. In running of the time N+I Atlanta' 99 matured his execution code and it created LDP-03. During N+I Atlanta' 99 went rewritable to follow the procedures of the appendix THE ONE of Lpd-05. In April of 2000 it continued the development of LDP to become portable. Then the package of the linux-mpls-ldp it was divided in two parts: mpls-linux (the plan base, seed of the forwarding of Linux liberated under GPL) and ldp-portable (a portable version of the protocol LDP liberated under LGPL that will be explained soon). The project was moved for Sourceforge in 30/11/2000. In January of 2003 the developers began to work for the iNOC (www.internetnoc.com) where they were motivated to work in this project and that you/they were involved in the standardization entities with MPLS.

\subsection{LDP-PORTABLE }

As informed in (RFC3036) the architecture MPLS defines a LDP (Protocol of Distribution of Labels) to dictate the procedures with the ones which LSR (Gateway and Switch of Labels) they inform to the other gateways as the labels will be used for sending and reception of traffic and their roads. 

\subsection{QUAGGA-MPLS} 
Quagga-mpls is a program to provide routing and implementations of OSPFv2, OSPFv3, RIPv1 
and v2,RIPv2 and BGPv4 in atmospheres Unix, particularly FreeBSD Linux and some NetBSD. 

\subsection{BGP/MPLS VPN }
As explained in RFC2547 is possible terms VPN distributed for several broken with the use of BGP, allowing that we have VPN 1 for N. 

\subsection{THE FERRAMENTA IPTRAF}
IPTraf is a generating program of net statistics for Linux based on console (character). It collects a variety of packages of the net such an as, the connection package TCP and the byte countings, the indicators of statistics and of traffic TCP/UDP for station of the LAN package and byte countings. These graphs aided us to analyze in real time and through logs the consumption and the direction of the traffic in each station. 

\subsection{THE FERRAMENTA IPERF}
Iperf is a software of band analysis and calculation of loss of datagrams in the net ipv4 or ipv6 that it is maintained by the University of Illinois under license GPL. [ Tirumala] 
We will be using the iperf with the keys-u (udp)-b defining the bandwidtch in values that vary from 10Mit/seg to 100Mbit/seg - t to define the time of transmission.
 
\subsection{THE FERRAMENTA TC}
TC (Traffic Controller) together with IPROUTE2 they are responsible for QOS in our experiments. This tool creates and it associates lines to the net output devices. She is used for several line parameters and to associate classes with those line parameters. TC can also be used to initialize filters with base in routing table, classifiers u32, classifiers tcindex and classifiers RSVP. [Almesberger].

\subsection{THE FERRAMENTA IPROUTE2 }
IPROUTE2 is a controller of calls of traffic. This application supports several classification methods, priorization , twist limitation and exit of I traffic. For these characteristics and for her lightness he will be used in ours analyze together with TC.

\subsection{THE FERRAMENTA IPBAND}
IPBAND is a monitor of traffic of IP. He divides the traffic in subnets and bands using logs to detail the band used in each sub-net. These data will be used to create the graphs for it analyzes. 

\subsection{THE FERRAMENTA NETPERF}
Netperf is a benchmark tool that can be used to measure different net types. He provides throughtput tests and latency point-to-point. The netperf provides tests in the atmospheres, TCP and UDP through BSD Sockets, DLPI, Unix Domain Sockets, Fore ATM API, HP HiPPi Link Level Access. (http://www.netperf.org/netperf/NetperfPage.html) 

\subsection{THE FERRAMENTA OPENVPN.}
Openvpn is a tool of creation of encrypted tunnels using SSL under the Internet. As great advantages are the portability among platforms, stability, support to dynamic IP and NAT among others. For to be of simple implementation and to use compacting we will use this tunnel type.
 
\subsection{THE FERRAMENTA PING}
Ping (Packet Internet Group) it is a program TCP/IP used to test the reach of a net, sending to us remote a request and waiting for an answer. 

\subsection{THE FERRAMENTA NS-2}
It is a simulator of net use. He has support for several simulations in TCP, routing and protocols multicast. ( Information Society Tecnologies IST - EuroNGI) 
\section{EXPERIMENTS ARE RESULTED }
\subsection{STRATEGY FOR THE TESTS }
In the accomplished rehearsals, the reservation of band passer-by in all the connections for the aggregation of flows EF is maintained at 220Kbps o'clock. This allocation represents 12\% of the connection of 2Mbps of the domain DiffServ. The remaining is destined to the traffic of better effort. Then we will have PHB EF priority 1 with 220 Kbps of band width and BE (Best Effort) with 1780Kbps. 
For the accomplished tests, four different cases were considered. First it was transmitted in the net only traffic of the type EF just considering the flow. The objective of that case is to investigate the behavior of the mechanisms of reservations of resources adopted by the tool TC. In the second case, it was transmitted in the net only traffic of the type EF, considering two flows. The objective here is the same of the previous case, added of the desire to investigate the behavior of a flow of the type EF, just considering a flow, and traffic type Better Effort (Best Effort-BE), with the purpose of investigating the behavior of the traffic EF in the presence of the traffic better effort. It is finally was transmitted in the net traffic of the type EF, now considering two flows, and better effort. Those last two cases and yours analyze characterize the main results of that work. See in the next section the analysis and comparison of the results. 

\section{ANALYSIS AND COMPARISON RESULTED DOS }
Starting from tests accomplished in the defined atmosphere, tests these that we will detail to proceed, were verified that, for the implementation of MPLS in Linux the actings of the technology MPLS were shown very superiors if compared to implementations IP, so much in the variation of delay, flow and tax of losses. It was observed that the routing IP had better acting just when used size packages small, smaller than 1024 bits, when the load of processing of LER's reduces the acting of MPLS. 

\subsection{Analyze of band passer-by}
With the use of the tools IPBand and Netperf made the tests of band passer-by in the net. Netperf was configured to send and to receive 20.000.000 transactions that, in spite of I number it, it lasts less than 2 seconds. During this period IPBand will be monitoring the sub-nets and generating reports informing the tax in Kbytes/100000.00 a second in each sub-net for the future generation of the graphs the they be analyzed. 

\subsection{Latency of Protocols.}
The main objective of this test type and to measure the times of pings latency in function of the I number of gateways that it crosses in her course, being compared among the two technologies, routing IP and MPLS. For this test bursts of 1000 ping were generated beginning with a gateway among two personal computers and finishing with nine gateways among two personal computers. We verified that using packages above 1024 bits not importing it numbers of gateways MPLS was shown faster in it processes it of packages, spending 33\% of the worn-out total time on average for the Roteamento IP for the same situation. 

\subsection{Tables of Routing / Labels }
Again we will use the ping as tool for us to measure the performance in our tests. In this step we will be creating tables of content routing 50000 broken. As it can be read in previous paragraphs MPLS ignores the tables making use of Labels that increases his/her performance when compared to the routing IP. We verified that MPLS was shown three times faster than the routing IP. Again the tests were made beginning with 3 conspire and ending with the job of 11 equipments being nine gateways MPLS Linux. 

\subsection{Load in Ping }
This test will show the benefits of the YOU (Traffic Engineering) that is one of the main differences between classic IP and administration of performance of MPLS as it can be seen in RFC2702 MPLS is strategically significant for the Engineering of Traffic he can potentiate provision of the functionality and readiness of the overlay model, with low cost front to other alternatives. Equally important he offers roads for LSP to pass. A LSP is a specification of the road for the keying Label. In practice the terms LSP and the Main traffic are used as synonyms. 
For this test we developed a script that will connect three points and it will saturate and to inform which the percentage of the band used during the test. It plans it 1 will accomplish one would be of 2000 pings. When it plans it 1 to finish accomplishing these pings, she will connect plans it then 2 anger to measure to the load in the circuit. When the traffic between 2 and 3 to arrive to the maximum value the test will be finished and it plans it 1 will generate a graph that will represent the evolution of the latency in function of the occupation. 
We saw that this test presents different results as we are going inserting other services on MPLS or the routing IP, being these layers the new gateways, the service of QOS and finally the installation of OpenVPN. In this I finish apprenticeship identified that MPLS added QOS and with the implementation of OpenVPN it presented indexes three smaller times when compared to the routing IP with QOS and OpenVPN.

\section{CONCLUSION}
The strong point in this study is the accomplishment of acting evaluations among the use of MPLS, added to the aspects that it supports like QOS and VPN verses routing added IP QOS and VPN. Both in atmosphere Linux, whose results are relevant, because most of the results found in our researches in the Internet are based in simulations or mathematical models or in versions previous and problematic of MPLS-LINUX. 
The conclusion that we arrived is that with the use of MPLS-LINUX added to tools of QOS and moving the data inside of tunnel VPN got to reach latency levels that make possible the replication of databases among servants with the use of DRBD and Heartbeat in a reliable way, it holds and with high performance inside of LAN and WAN. Traffics among servants in a net LAN wins up to three times more speed when submitted to this cocktail (MPLS/QOS/VPN) when compared to routing IP with QOS and VPN in nets LAN.

\end{document}